# Electrocapillary instability of magnetic fluid peak

*Levon Mkrtchyan, Arthur Zakinyan,\* Yuri Dikansky*

Department of Physics, Institute of Mathematics and Natural Sciences, North-Caucasian Federal University, 1 Pushkin Str., 355009 Stavropol, Russian Federation

ABSTRACT: The paper presents an experimental study of the capillary electrostatic instability occurring under effect of a constant electric field on a magnetic fluid individual peak. The peaks under study occur at disintegration of a magnetic fluid layer applied on a flat electrode surface under effect of a perpendicular magnetic field. The electrocapillary instability shows itself as an emission of charged drops jets from the peak point in direction of the opposing electrode. The charged drops emission repeats periodically and results in the peak shape pulsations. It is shown that a magnetic field affects the electrocapillary instability occurrence regularities and can stimulate its development. The critical electric and magnetic field strengths at which the instability occurs have been measured; their dependence on the peak size is shown. The hysteresis in the system has been studied; it consists in that the charged drops emission stops at a lesser electric (or magnetic) field strength than that of the initial occurrence. The peak pulsations frequency depending on the magnetic and electric field strengths and on the peak size has been measured.

INTRODUCTION

The fluid surface capillary electrostatic instability development and the electrostatic dispersion traditionally attract attention of researchers. This is due to varied scientific and practical applications of these in the applied physics, geophysics, scientific tools engineering, etc (see, e.g., [1,2] and the literature there). Thus, among other things, the studies of the storm events and various occurrences of the atmospheric electricity make necessary the studies of the freely suspended fluid drops instability and dispersion [3]. Lot of works deal with the experimental and theoretical studies of the disintegration of a charged drop instable in relation to its own charge [4,5], of an uncharged drop instable in relation to polarization charge in a strong electrostatic field [6–10], and, in general, of a charged drop in an electric field [1,3,11].

The fluid electrostatic spraying is widely used in the aerosol sprays technology [12]; it yields better results than the air spray methods. The electrostatic spraying based mass spectrometry methods are widely used, as well [13]. These methods are, as a rule, based on the effect of the finely divided drops jets emission from the end of a charged cylindrical capillary connected to a fluid tank at a certain distance from the flat opposing electrode [14–17].

The individual patterns of the electrocapillary instability development may differ depending on the experimental conditions; however, in any event, they have a number of common features and regularities consisting in the following. When a fluid free surface gets a charge or is exposed to an external electric field, it deforms, usually into a spheroid shape that transforms later on into a pointed cone shaped configuration. Such behavior results from the electric and surface tension forces competition. When the intrinsic or polarization charge on a fluid surface comes to critical, the excess of charge discharges in form of charged drops emission from the cone point. The charged drops emission may periodically repeat resulting in the fluid surface pulsation [15–17]. The drops periodical emission under effect of electric field is applied in the hi-res ink jet printing [17].

Recently, the capillary electrostatic instability studies have regained momentum due to the development of the microfluidic technologies and creation of the so called "labs-on-a-chip" that widely use the small fluid volumes control by external fields. Such applications have interest in the studies of the electrostatic stability of the drops on a solid conducting surface in an electric field [18–21]. This problem is also linked with a number of classic problems, such as higher losses in power lines in rainy weather, studies of the St. Elmo's light (they occur due to the electrostatic instability of the water drops depositing on high objects) occurrence regularities, etc.

Despite the multitude of applications and large number of experimental and theoretical studies, certain aspects of the fluid electrocapillary instability occurrence remain insufficiently known, e.g., the instability development time behavior. The new research capacities and potential practical applications emerge due to the magnetic fluids use in studying the electrocapillary instability. A magnetic fluid is a stable colloidal dispersion of magnetic nanoparticles in a carrier fluid [22, 23]. It can intensively interact with a magnetic field, and may, in the framework of the studies presented here, be considered as a fluid magnetizing medium. Its capacity to interact with both the electric and magnetic fields makes it a unique medium. The additional exposure to a magnetic field gives the possibility to act on the development process of the electrostatic instability of magnetic fluid by changing its free surface shape; hence, the new possibilities in studying the instability regularities and in improving a number of the electric spraying engineering applications appear.

The behavior of a magnetic fluid drop freely suspended in a surrounding nonmagnetic liquid and exposed to a magnetic field was studied experimentally and theoretically in many works (see [23] for a review). Investigation of the influence of bounding surfaces on the deformation and dynamics of a drop is of considerable interest from the viewpoint of potential applications for the microfluidic devices. Such investigations for a drop of a magnetic fluid are presented in works [24–30], where the behavior of the drops lying on a solid surface and surrounded by air has been studied.

The behavior of magnetic fluid drops subjected to simultaneously acting electric and magnetic fields still remains relatively unstudied. The study of the simultaneous effect of an electric and a magnetic field on a stationary shape of a magnetic fluid drop is presented in works [31–33]. The electrocapillary instability of a magnetic fluid was not investigated previously. This article deals with the specificities of capillary electrostatic instability of a magnetic fluid microdrop placed on a flat electrode surface in electric field under additional effect of external magnetic field.

EXPERIMENTAL DETAILS AND RESULTS

Figure 1 shows the experimental setup layout. It consists of a horizontal flat capacitor connected to a constant voltage source. The 1.5×2.5 cm rectangular shape capacitor sheets are made of glass coated with thin transparent electrically conductive stannic dioxide. The distance between electrodes is 3.5 mm. A thin (~ 10 μm) magnetic fluid layer was applied on the capacitor lower sheet, on the inside. The magnetic fluid used in experiments was a magnetite nanoparticles (~ 10 nm) dispersion in kerosene; the oleic acid was used as stabilizer. The magnetic fluid density was 1400 kg/m$^3$, the initial magnetic permeability was 6.3, the magnetic nanoparticles concentration was 13%, the saturation magnetization was 55.4 kA/m, the dielectric permeability was 5.2, the specific electric conductivity was $1.3 \cdot 10^{-6}$ S/m, the air interfacial tension was 0.028 N/m, and the dynamic viscosity was 30 mPa·s.

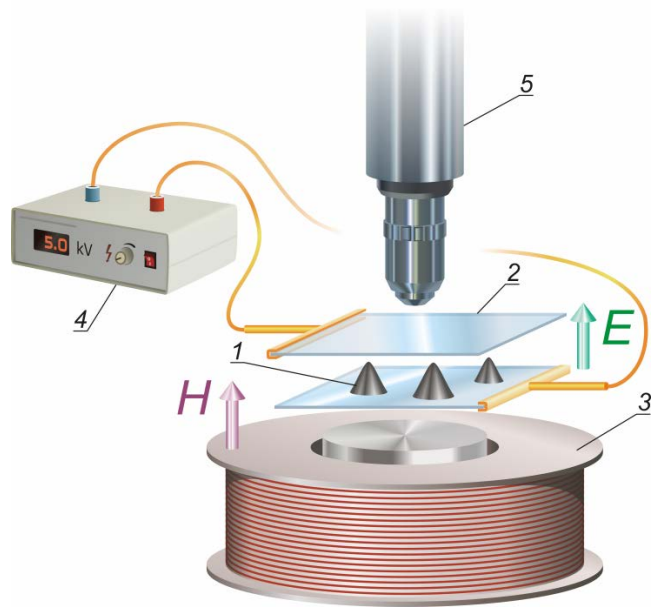

**Figure 1.** Layout of experimental setup to study magnetic fluid peak electrocapillary instability in magnetic and electric fields: 1 – magnetic fluid peaks, 2 – capacitor electrodes, 3 – electromagnet, 4 – high voltage source, and 5 – microscope.

The capacitor was installed onto the pole piece of an electromagnet creating a vertical magnetic field. After activation of the magnetic field whose intensity was above a certain critical value, the development of the magnetostatic instability of the surface of the magnetic fluid layer applied on the lower electrode began resulting in the layer disintegration into separate cone shaped drops (peaks). This phenomenon has been studied in a number of works (see, e.g., [22]). The peaks were oriented along the magnetic field; their shape depended on the magnetic field strength and magnetic fluid layer initial thickness. The thicker was the magnetic fluid layer, the greater were the peaks resulting from its disintegration in magnetic field. The above procedure allowed for the small size magnetic fluid drops, impossible with other methods. The magnetic field within the limits of the small region of space occupied by the specimen under study could be considered uniform; its intensity $H$ varied between 0 to 100 kA/m. With increasing magnetic field intensity the peak elongates in the direction of the applied field, remaining axisymmetric. The behavior of a sessile magnetic fluid droplet under the action of magnetic field is reviewed in [22, 23]. Then a constant electric voltage was applied to the capacitor electrodes, resulting in the electric field of the same direction as the magnetic one. The range of applied voltages $U$ was 0 to 6 kV. The magnetic fluid behavior was studied under an optical microscope. The observations have been made from above for viewing the events along the electric and magnetic fields direction and from one side for viewing perpendicularly to the fields direction. As the quantitative measurements were made by observation from above, in reflected light, only the peak base size could be determined. Further on, the base size of a peak resulting from the magnetic fluid layer exposure to magnetic field will be used to characterize the peaks sizes. The experimental peaks base sizes were 400 to 900 μm.

It was found that a magnetic fluid peak exposed to an electric field extends along the field and the peak's base visible diameter reduces. When reaching a certain critical applied voltage periodical peaks shapes pulsations were observed. The peak shape dynamics was shot on a high speed digital video camera. It has been revealed that the peaks shape pulsation occurs due to the electrocapillary instability final phase development; the pulsation is accompanied by an emission of finely divided charged drops jets from the peak point in direction of the opposing electrode. The sizes of the emitted droplets are of the order of 1 μm. After such emission a peak loses some of its charge, hence, its shape changes under effect of the capillary and gravity forces; then the charge accumulates anew, the peak stretches along the field, and the new charged drops jets emission occurs. This process repeats with a clearly traceable periodicity. Figure 2 shows a number of photo shots demonstrating the peak shape pulsation at charged drops jets emission from its point. The video clip enclosed with this article illustrates the peak pulsations process decelerated by a factor of 20, as well. As is seen, at first the magnetic fluid peak shrinks. Then the peak stretches at the jet emission. The described process regularities were independent of the capacitor electrodes polarities. It should be noted that in experimental conditions certain magnetic fluid peaks were far enough from each other (the distances were significantly greater than the peaks sizes); hence, the interaction between peaks effect on the processes under study can be neglected.

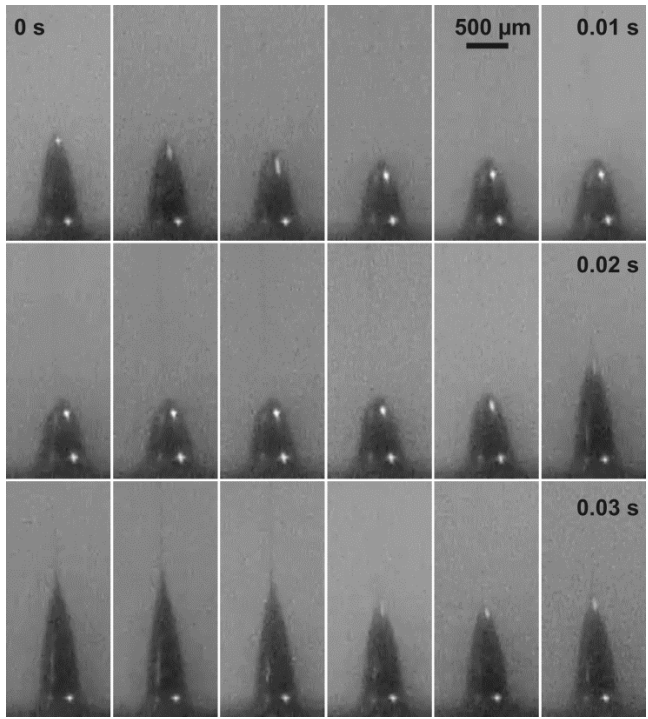

**Figure 2.** Progressive photo shots of magnetic fluid pulsating peak (one pulsation cycle; side view) at 1/600 s time intervals (at $U = 1.5$ kV; $H = 31$ kA/m). For the complete observation of the peak behavior see movie.

The critical voltage $U_c$ at which the charged drops emission from the peak point occurs was measured. The critical voltage turned out to be dependent on the peaks sizes. Figure 3 shows the experimental dependence of the critical electric voltage on the peak base diameter $d$. It should be pointed out that here and in what follows the base diameter means the base diameter at the peak formation moment and without electric field. As Figure 3 shows, the critical voltage decreases as the peaks size grows. The electrocapillary instability occurrence must also depend on the magnetic field acting on the peak. Indeed, the instability occurs when the charge density on peak point reaches critical value; hence, acting on the peak shape with a magnetic field allows changing the electric charge density on the point. Thus, in particular, the magnetic field intensity growth results in a lesser angle at the peak point: this must stimulate the electrocapillary instability development. The studies have shown that fixing the electric voltage at the level where there is no charged droplets emission and increasing the magnetic field intensity may also lead to the electrocapillary instability when the magnetic field intensity reaches a certain critical value. Figure 3 shows as an example the experimental magnetic field critical intensity $H_c$ dependence on the peak base diameter; at such intensity the peak shape pulsations occur. It can be seen that this dependence monotonely decreases.

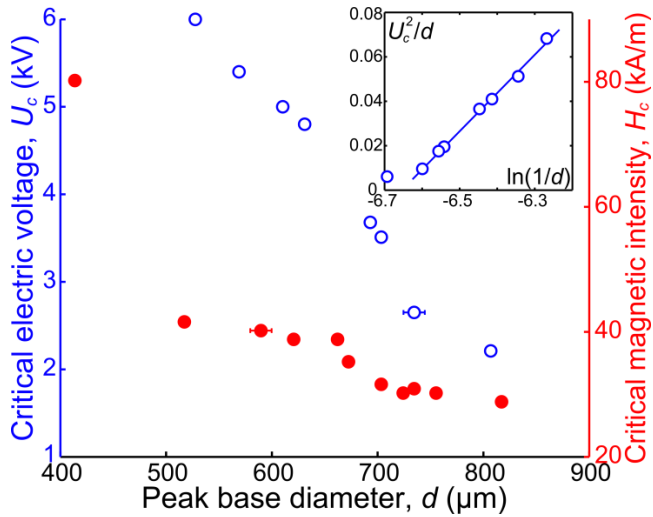

**Figure 3.** Experimental dependences of critical electric voltage (○, at $H = 27$ kA/m) and critical magnetic field intensity (●, at $U = 3$ kV) at which electrocapillary instability occurs on peak base diameter.

Obviously, the critical electric and magnetic fields at which the final phase of the peak electrocapillary instability occurs must be interdependent. Figure 4 shows the experimental dependence of the critical electric voltage on the critical magnetic field intensity. This dependence was measured as follows. First, certain magnetic field intensity was reached, and then the electric voltage was applied and gradually increased up to the moment when the fluid jets emission from the peak point began; the procedure was repeated for another magnetic field intensity. As Figure 4 shows, the critical electric voltage dependence on the critical magnetic field intensity monotonely decreases. If, after the magnetic fluid peak pulsations start, the electric (or magnetic) field begins reducing, the pulsation stops at a lesser electric (or magnetic) field than the starting one. Figure 4 also shows the dependence of the critical electric voltage at which the electrocapillary instability disappears on the critical magnetic field intensity; this demonstrates the hysteresis in the system under study.

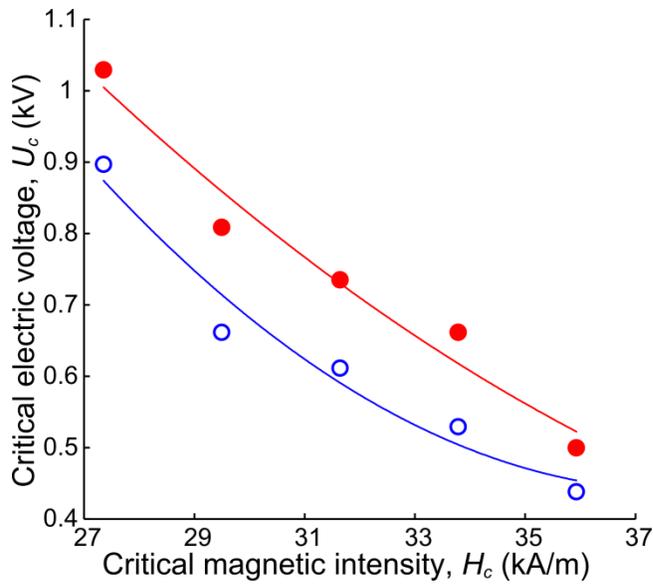

**Figure 4.** Critical electric voltage and magnetic field intensity corresponding to start (●) and stop (○) of electrocapillary instability (at $d = 765$ μm). Lines are the approximation of experimental data.

The magnetic fluid peak pulsations frequency $f$ was measured. It has turned out that the peak pulsations frequency grows with the electric voltage and magnetic field intensity growth. Figure 5 shows the relevant experimental dependences. Figure 6 shows the experimental dependence of the pulsations frequency on the peak base diameter: with the peaks sizes growth their pulsations frequency decreases.

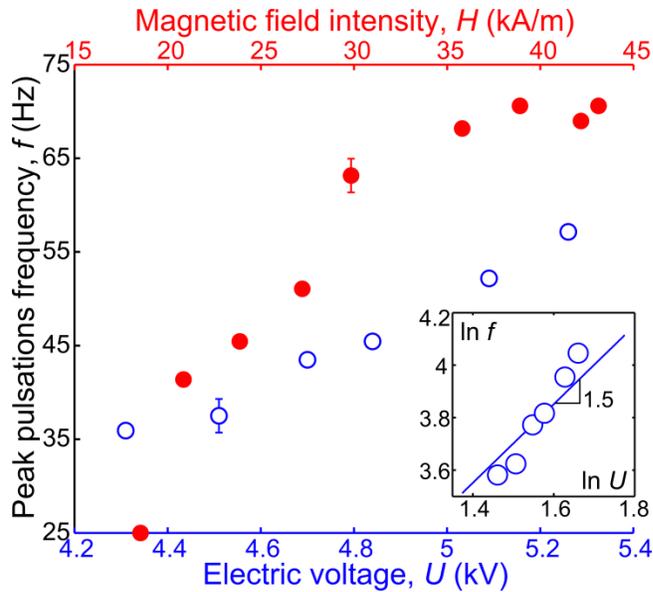

**Figure 5.** Experimental dependences of peak pulsations frequency on electric voltage (○, at $H = 27$ kA/m, $d = 630$ μm) and on external magnetic field intensity (●, at $U = 4.1$ kV, $d=775$ μm).

When measuring the peak pulsations frequency, every frequency was measured at least 20 times. The frequency determination error was then calculated using the standard multiple measurements statistical errors determination methodology, at the 0.9 confidence level. The magnetic field intensity measurement error was determined by the instrumental error of the current meter used to control the electromagnet coil current. The electric voltage measurement error was determined by the instrumental error of the voltmeter. The absolute values of these errors do not exceed the markers on the experimental curves. The peak base diameter measurement error was ~ 10 μm; it was conditioned by a certain peak lower bound image blur.

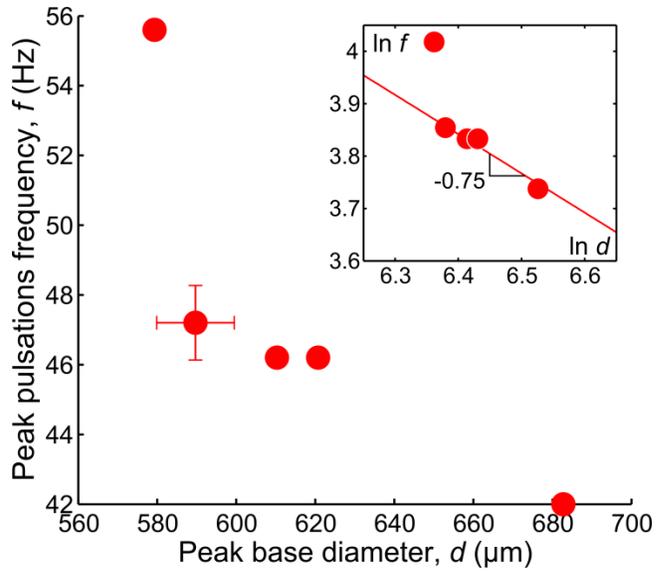

**Figure 6.** Experimental dependence of peak pulsations frequency on its base diameter (at *H* = 27 kA/m, *U* = 5 kV).

It was observed that under the action of comparatively strong electric and magnetic fields the stretched magnetic fluid peak can reach the opposite (upper) electrode and form the column between electrodes. The shape of such column vibrates. As an example, a series of snapshots of the vibrating column made in a vertical plane are shown in Figure 7.

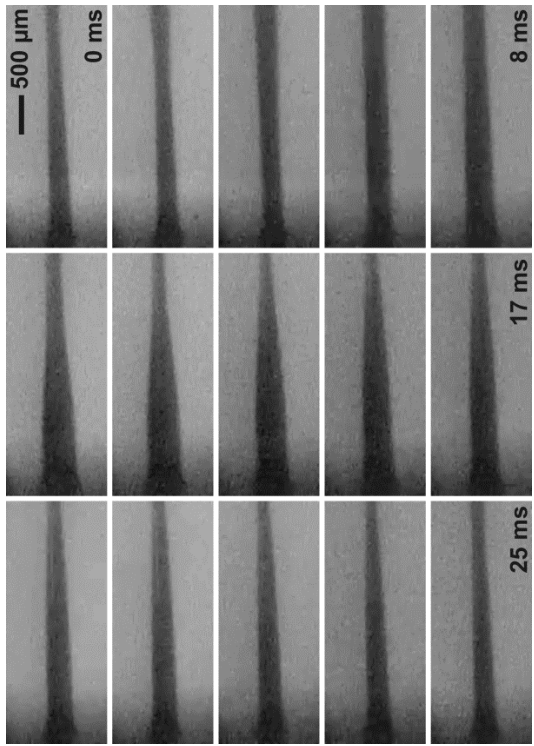

**Figure 7.** Consecutive snapshots at 1/600 s time intervals showing a magnetic fluid pulsating column between the capacitor electrodes (one pulsation cycle is shown; at $U = 3.1$ kV; $H = 60$ kA/m).

In conclusion, it should be noted that the electric field effect in the experimental voltages range never led to any significant effect on the structure, geometry, or pattern of decomposition in magnetic field of the initial magnetic fluid layer. The electric field effect was visible only for the magnetic fluid peaks resulting from the layer decomposition in the presence of magnetic field.

ANALYSIS AND DISCUSSION OF RESULTS

As noted above, the revealed magnetic fluid peak shape pulsations result from the development of the electrocapillary instability final phase. For the periodical emission of the charged droplets jets from the peak point to result in the visible fluctuations of its shape, the time of the final instability phase development must be greater than the peak shape relaxation characteristic time. In such case the peak shape will have time to change between the excess charge emission moments; this is visible in the peak pulsations form. The peak shape relaxation time may be estimated from the expression $\tau \approx \eta R/\sigma$, where: $\eta$ is the fluid dynamic viscosity, $R$ the peak characteristic size, $\sigma$ the surface tension. Substituting the relevant parameters values get that the peak shape relaxation time is, indeed, by two orders lesser than that of its pulsations period; hence, the described pulsations mechanism is correct.

The jetting at the apex of the peak can be explained as a competition between surface tension and electric field forces. When the electric field exceeds a certain critical value, the jetting begins at the tip of the conical magnetic fluid peak. It is customary to define the dimensionless numbers characterizing the relative ratio between the forces having a profound impact in the experiment. The Bond number is a ratio between the surface tension and the gravity forces Bo, the magnetic Bond number is a ratio between the surface tension and the magnetic field forces $Bo_m$, the ratio between the surface tension and the electric field forces is characterized by electrical capillary numbers $C_E$ and $C_Q$:

$$\mathrm{Bo} = \frac{\rho g R^2}{2\sigma}; \quad \mathrm{Bo}_m = \frac{\mu_0 H^2 R}{4\sigma};$$

$$C_E = \frac{\varepsilon_0 E_0^2 R}{4\sigma}; \quad C_Q = \frac{3Q^2}{32\pi^2 \varepsilon_0 \sigma R^3},$$

where: $\rho$ is the magnetic fluid density; $g$ the acceleration of gravity; $\mu_0 = 4\pi \cdot 10^{-7}$ H/m is the permeability of vacuum; $\varepsilon_0 = 8.85 \cdot 10^{-12}$ F/m is the permittivity of free space; $E_0$ is the electric field intensity, which can be estimated as $E_0 = U/h$ (here $h$ is the distance between electrodes); $Q$ is the charge on the magnetic fluid peak. As it was shown in [34], the charge of the conducting droplet contacting with an electrode can be calculated by following expression $Q \approx 20.68 \cdot \varepsilon_0 E_0 R^2$. Within the range of the experimental parameters values we obtain that $Bo_m > C_Q > 1$ and $C_E < Bo < 1$. It can be concluded that in the above-described experiments the magnetic field and the electric charge of a magnetic fluid peak play key role in the peak shape deformation and instability development.

Consider the electrocapillary instability occurrence conditions, in particular, the critical electric voltage at which the charged droplets emission starts. In [14], the charged drops emission from the end of a capillar with fluid was studied. The capillar was standing vertically on the capacitor lower electrode, in the middle of the interelectrode space. This situation resembles the one studied here; the difference is that in our case the jets emission comes not from a cylindrical capillar but from a cone shaped fluid peak. For the critical voltage at which the emission occurs the following expression was obtained in [14]:

$$U_c^2 = \frac{2h^2}{\varepsilon_0 l^2}\left[\ln\left(\frac{4l}{d}\right) - \frac{3}{2}\right] \cdot 1.3\pi d\sigma, \qquad (1)$$

where: $l$ the capillar height, $h$ the distance between capacitor electrodes, and $d$ the capillar diameter. Obviously, the quantitative similarity between the experimental results of this work and expression (1) cannot be expected; however, these results may be compared qualitatively. Consider the critical voltage dependence on peak size. In expression (1), $d$ will mean the peak base diameter. According to (1), the $U_c^2/d$ dependence on $\ln(1/d)$ will be linear. Figure 3 shows the curve of the relevant experimental dependence in the stated coordinates axes; it shows that the dependence is, indeed, linear.

Consider the peak shape pulsations frequency. In [17] for the frequency of the charged droplets emission from the end of cylindrical capillar the following expression was obtained:

$$f \propto \left(\frac{\varepsilon_0^3}{\rho^2 \sigma}\right)^{1/4} \frac{E^{3/2}}{d^{3/4}}, \qquad (2)$$

where: $\rho$ is the fluid density, $E$ the electric field intensity at the end of capillar. Consider the intensity $E$ directly proportional to the applied voltage $U$. According to (2), the dependence of $\ln(f)$ on $\ln(U)$ must be linear with 1.5 slope ratio. Figure 5 shows the curve of the relevant experimental dependence approximated by a linear function with a 1.5 slope ratio; it is seen that the experimental data character agrees with the one predicted by expression (2). According to (2), the dependence of $\ln(f)$ on $\ln(d)$ must be linear with –0.75 slope ratio. Figure 6 shows the curve of the relevant experimental dependence approximated by a linear function with a –0.75 slope ratio; it is seen that the experimental data character agrees well enough with the one predicted by expression (2) in this case, as well. These results mean that the physical mechanisms of the electrocapillary instability dealt with here are similar to the ones of [14, 17].

It follows from the experimental data that the additional exposure to a magnetic field stimulates the electrocapillary instability development. This must be due, as already mentioned, to the magnetic field effect on the magnetic fluid peak shape. Thus, in particular, magnetic field growth results in a smaller peak point curvature radius. As a result, the electric charge density at the point grows; hence, the earlier occurrence of the instability final phase and charged drops jets emission. This is also why the magnetic field growth results in the peak shape pulsations frequency growth: the instability development requires here a lesser charge accumulation taking less time. It should be, however, noted that a more detailed quantitative analysis of the magnetic field effect on the magnetic fluid peak electrocapillary instability is difficult from the mathematical point of view and will not be performed here. It should be also noted that the revealed hysteresis in the electrocapillary instability development cannot be, either, explained within the available theoretical models framework and requires additional research.

CONCLUSION

The experimental studies have shown that a magnetic field can be used to work on the magnetic fluid electrocapillary instability development regularities. It has been specifically shown that a magnetic field codirectional with an electric one stimulates the magnetic fluid peak instability development and such instability occurs at smaller electric fields than without magnetic field. This may be useful when upgrading the electrostatic dispersion methods. The revealed magnetic fluid peak capillary electrostatic instability development regularities are, in whole, explainable within the available models framework; however, additional research is required to get a strict theoretical description of the magnetic fluid effect on the instability character and to analyze the revealed hysteresis.

ASSOCIATED CONTENT

**Supporting Information**.

Video clip illustrating the peak pulsations process decelerated by a factor of 20. This material is available free of charge via the Internet at http://pubs.acs.org.

AUTHOR INFORMATION

**Corresponding Author**

* E-mail zakinyan.a.r@mail.ru.

**Notes**

The authors declare no competing financial interest.

ACKNOWLEDGMENTS

This work was supported by the grant of the President of the Russian Federation No. MK-6053.2012.2 and by the Ministry of Education and Science of the Russian Federation in the framework of the scientific program "Development of Scientific Potential of Higher School".

(34) Shutov, A.A. Charge and force acting on a spherical body near a planar electrode in a polarizable conducting medium. *Tech. Phys*. **2003**, *48*, 669–676.

Table of Contents graphic

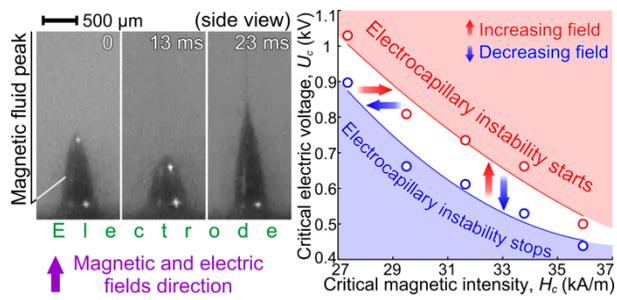